\documentclass[11pt,preprint]{aastex}
\usepackage{bm}
\usepackage{epsfig}

\begin{document}

\title{\bf Magnetised cosmological perturbations in the post-recombination era}

\author{Hera Vasileiou\footnote{Current address: Department of Physics, University of Crete, Heraklion 70013, Greece}~ and Christos G. Tsagas}

\affil{Section of Astrophysics, Astronomy and Mechanics, Department of Physics,\\ Aristotle University of Thessaloniki, Thessaloniki 54124, Greece}

\date{\empty}

\begin{abstract}
We study inhomogeneous magnetised cosmologies through the post-recombination era in the framework of Newtonian gravity and the ideal-magnetohydrodynamic limit. The nonlinear kinematic and dynamic equations are derived and linearised around the Newtonian counterpart of the Einstein-de Sitter universe. This allows for a direct comparison with the earlier relativistic treatments of the issue. Focusing on the evolution of linear density perturbations, we provide new analytic solutions which include the effects of the magnetic pressure as well as those of the field's tension. We confirm that the pressure of field inhibits the growth of density distortions and can induce a purely magnetic Jeans length. On scales larger than the aforementioned characteristic length the inhomogeneities grow, though slower than in non-magnetised universes. Wavelengths smaller than the magnetic Jeans length typically oscillate with decreasing amplitude. We also identify a narrow range of scales, just below the Jeans length, where the perturbations exhibit a slower power-law decay. In all cases, the effect of the field is proportional to its strength and increases as we move to progressively smaller lengths.
\end{abstract}

\section{Introduction}\label{sI}
The origin and the potential implications of the large-scale magnetic fields seen in the universe today remain an open challenge. Most structure formation models, the $\Lambda$CDM scenario in particular, bypass these fields despite their widespread presence~\citep{K,CT,HW,B,V}. The fact that the galactic, the cluster and the protogalactic magnetic fields have comparable strengths, between $10^{-7}$ and $10^{-6}$ Gauss, could be interpreted as a sign that cosmic magnetism has cosmological origin, but the issue is far from closed~\citep{GR,Wi,KKT,Retal}. Cosmological magnetic ($B$) fields could have played a role during both the early and the late stages of structure formation, since they introduce new ingredients to their environment. During the dust epoch, for example, large-scale $B$-fields (if present) are essentially the sole source of pressure support. The latter is known to inhibit the growth of density perturbations and also determine the scale of the Jeans length, which in turn decides the size of the first structures to collapse gravitationally. An additional, rather unique, magnetic property is their tension, reflecting the elasticity of the field lines and their tendency to react against any agent that distorts them.

In the present article we analyse the effects of large-scale magnetic fields on the evolution of baryonic density inhomogeneities after recombination and until the moment the recent accelerated expansion of the universe starts. We do so by confining to subhorizon scales and use the Newtonian version of the 1+3 covariant approach to cosmological perturbation theory~\citep{EvE,TCM}. After a brief introduction to the formalism, we proceed to analyse the magnetic component within the framework of the ideal magnetohydrodynamics (MHD). The latter is believed to provide an accurate approximation of the observable universe, which appears to be a very good electrical conductor. We follow the steps of earlier relativistic treatments~\citep{TB1,TB2,TM,BMT}, which also allows for a direct comparison between the two approaches. At the same time, the relative simplicity of the Newtonian analysis enables a closer and more detailed look into the magnetic effects on density perturbations after recombination. We begin by providing the nonlinear expressions monitoring the evolution of inhomogeneities in the density distribution of the (highly conductive) baryonic matter. These are subsequently linearised around an Einstein-de Sitter background that also contains a weak and fully random magnetic field. Focusing on the role and the implications of the magnetic pressure, we confirm that it establishes a purely magnetic Jeans length, the scale of which depends on the field's strength. We also provide a new (to the best of our knowledge) set of analytic solutions, revealing that the magnetic effect on linear density inhomogeneities is scale-dependent. More specifically, perturbations larger than the aforementioned Jeans length grow, but at a pace slower than in non-magnetised models. Wavelengths smaller than the Jeans length, on the other hand, typically oscillate with a time-decaying amplitude. Finally, we find that there is a narrow margin of scales, just inside the Jeans length, where the density inhomogeneities experience a (relatively) slow power-law decay. In all cases, the magnetic presence inhibits the growth of density perturbations, but the exact nature of the effect depends on the field's strength and on the scale under consideration.

Adopting the Newtonian framework also allowed us to incorporate the effects of the magnetic tension into the linear equations  (see Appendix~A). As we mentioned above, these effects carry the reaction of the field lines to distortions caused by changes in their shape and their orientation. Nevertheless, at the linear perturbative level, the role of the magnetic tension seems secondary compared to that of the field's pressure. Finally, for completeness and also for comparison, we consider and discuss the magnetic effects on linear density perturbations assuming a pure (i.e.~magnetic-free) Einstein-de Sitter background (see Appendix~B).

\section{Newtonian covariant
magnetohydrodynamics}\label{sNCMHD}
The covariant formalism (see~\cite{EvE,TCM} for reviews) has been applied to cosmology by many authors. Electromagnetic fields have been studied using covariant techniques in~\cite{Eh,E2,T,ST} and the formalism was applied to magnetised cosmological perturbations in~\cite{TB1} (see also~\cite{BMT} for a review). Nevertheless, a Newtonian covariant study of perturbed magnetised cosmologies has been missing from the literature.

\subsection{Kinematics}\label{ssCs}
We use general coordinates $\{x^a\}$, with $a=1,2,3$, to define the metric tensor ($h_{ab}$) of the Euclidean space. The metric and its inverse matrix ($h^{ab}$) satisfy the constraints $h_{ac}h^c{}_b= \delta_{ab}$, with $\delta_{ab}$ being the Kronecker symbol. In a Cartesian frame, where $h_{ab}= \delta_{ab}$, the covariant and contravariant components coincide and spatial derivatives are replaced by ordinary partial derivatives (e.g.~see~\cite{E1,E2}). We adopt a fluid approximation, where the matter and the comoving (fundamental) observers are moving with velocity $v_a$ tangent to the flow lines. The `convective' derivative of a given tensorial quantity $T$ is therefore given by $\dot{T}=\partial_tT+ v^a\partial_aT$, where the first term on the right-hand side describes changes at a fixed point and the second takes into account the motion of the fluid. Hence, the convective derivative of the fluid velocity,
\begin{equation}
\dot{v}_a= \partial_tv_a+ v^b\partial_bv_a\,,  \label{dotva1}
\end{equation}
provides the inertial acceleration.

The complete kinematic description of the fluid, as well as that of the associated comoving observers, follows from the spatial gradient of the velocity field. Like any second rank tensor, the derivative $\partial_bv_a$ decomposes as
\begin{equation}
\partial_bv_a= {1\over3}\,\Theta\delta_{ab}+ \sigma_{ab}+ \omega_{ab}\,,  \label{pbva}
\end{equation}
where $\Theta=\partial^av_a$, $\sigma_{ab}=\partial_{\langle b} v_{a\rangle}$ and $\omega_{ab}=\partial_{[b}v_{a]}$.\footnote{Round brackets in the indices denote symmetrisation, square indicate antisymmetrisation and angled ones define the symmetric and trace-free part of second-rank tensors. Therefore, $\partial_{\langle b}v_{a\rangle}= \partial_{(b}v_{a)}- (\partial^cv_c/3)\delta_{ab}$.} The scalar $\Theta$ describes variations in the fluid volume, while the tensors $\sigma_{ab}$ (shear) and $\omega_{ab}$ (vorticity) monitor changes in the shape (under constant volume) and in the orientation (i.e.~rotation) respectively. Positive values of $\Theta$ indicate expansion and negative contraction. The volume scalar also defines a representative length scale ($a$) according to $\dot{a}/a= \Theta/3$. In cosmological studies, the aforementioned length coincides with the scale factor of the universe.

\subsection{Hydrodynamics}\label{ssHDs}
Within the Newtonian framework, the gravitational field is described by the potential ($\Phi$), which obeys the Poisson equation
\begin{equation}
\partial^2\Phi= {1\over2}\,\kappa\rho- \Lambda\,,
\label{P}
\end{equation}
where $\partial^2=\partial^a\partial_a$ is the Laplacian, $\rho$ is the matter density and $\kappa=8\pi G$. Also, the cosmological constant ($\Lambda$) is measured in units of inverse-time squared. Given that $\partial_a\Phi$ is the gravitational acceleration and $\dot{v}_a$ its inertial counterpart (see Eq.~(\ref{dotva1})), the vector
\begin{equation}
A_a= \dot{v}_a+ \partial_a\Phi\,,  \label{Aa}
\end{equation}
carries the combined action of gravitational and inertial forces. Therefore, $A_a$ is the Newtonian analogue of the relativistic 4-acceleration vector~\citep{E1,E2}. The density and the total acceleration of the fluid satisfy their conservation laws, namely the continuity equation and the Navier-Stokes formula, which take the covariant form
\begin{equation}
\dot{\rho}=-\Theta\rho \hspace{15mm} {\rm and} \hspace{15mm}
\rho A_a= -\partial_ap- \partial^b\pi_{ab}\,,  \label{CEs}
\end{equation}
respectively. Here, $p$ is the isotropic and $\pi_{ab}$ is the anisotropic pressure of the medium (with $\pi_{ab}=\pi_{\langle ab\rangle}$). Finally, one also needs an equation of state relating the pressure to the density. Since we will be dealing with non-relativistic matter, the pressure is thermal in nature and the associated sound speed is very small (compared to that of light, namely $c_s^2={\rm d}p/{\rm d}\rho\ll1$).

To complete the hydrodynamic description, we need the propagation equations of the three kinematic variables and an equal number of constraints. For our purposes, the key formula is the propagation equation of the volume scalar (see~\cite{E1} and~\cite{ST} for more details), namely the Newtonian version of Raychaudhuri's equation,
\begin{equation}
\dot{\Theta}= -{1\over3}\,\Theta^2- {1\over2}\,\kappa\rho+ \partial^aA_a- 2(\sigma^2-\omega^2)+ \Lambda\,,  \label{Ray}
\end{equation}
with $\sigma^2=\sigma_{ab}\sigma^{ab}/2$ and $\omega^2= \omega_{ab}\omega^{ab}/2$ giving the shear and the vorticity magnitudes respectively.

\subsection{Ideal magnetohydrodynamics}\label{ssIMHD}
In an electrically conducting medium, the charges are carried by the positive ions and by the electrons. Assuming global electrical neutrality and adopting a single-fluid approach, the velocity of the matter is that of the centre of mass (e.g.~see~\cite{G,ST})
\begin{equation}
v_a={1\over m_++m_-}\,(m_+v_a^++m_-v_a^-)\,,  \label{cmvel}
\end{equation}
where $m_{\pm}$ are the masses of the ions and the electrons and $v_a^{\pm}$ their velocities. Then, at the ideal-MHD limit the electrical conductivity diverges and Ohm's law reduces to a simple relation between the electric ($E_a$) and the magnetic field ($B_a$).
\begin{equation}
E_a= -\epsilon_{abc}v^bB^c\,,  \label{MHDOhm}
\end{equation}
where $\epsilon_{abc}$ is the Levi-Civita tensor. On using the above, the magnetic induction equation takes the covariant form (see~\cite{ST} for the details)
\begin{equation}
\dot{B}_a= -{2\over3}\,\Theta B_a+ (\sigma_{ab}+\omega_{ab})B^b\,,  \label{MHDM1}
\end{equation}
guaranteeing that the magnetic vector connects the same particles at all times~\citep{E2}. In other words, when the ideal-MHD limit holds, the $B$-field is frozen into the matter.

The magnetic field is a source of isotropic and anisotropic pressure (e.g.~see~\cite{P,M}), with both contributing to the Lorentz force. At the ideal-MHD limit, the latter recasts the Navier-Stokes equation into~\citep{ST}
\begin{equation}
\rho A_a= -\partial_ap- \partial^b\pi_{ab}- {1\over2}\,\partial_aB^2+ B^b\partial_bB_a\,,  \label{NSMHD3}
\end{equation}
which replaces the standard expression given in Eq.~(\ref{CEs}b). Note that the third term on the right-hand side is due to the magnetic pressure and the fourth comes from the field's tension. When these two stresses balance each other out, the Lorentz force vanishes and the $B$-field can reach equilibrium. The effect of the tension term vanishes when the field lines are `geodesics', in which case $B^b\partial_bB_a=0$.

In the ideal-MHD limit, the matter and the $B$-field are conserved separately. In particular, the fluid continuity equation retains its `non-magnetised' form (see Eq.~(\ref{CEs}a) earlier). On the other hand, contracting the induction equation (see expression (\ref{MHDM1})) along the field vector gives
\begin{equation}
\left(B^2\right)^{\cdot}= -{4\over3}\,\Theta B^2- 2\sigma_{ab}\Pi^{ab}\,,  \label{MHDdotB2}
\end{equation}
where $\Pi_{ab}=(B^3/3)\delta_{ab}-B_aB_b$. The above can be seen as the conservation law of the magnetic isotropic pressure~\citep{ST}. Similarly, starting again from Eq.~(\ref{MHDM1}), one obtains the conservation law of the field's anisotropic pressure, namely
\begin{equation}
\dot{\Pi}_{ab}= -{4\over3}\,\Theta\Pi_{ab}+ 2\Pi_{c\langle
a}\sigma^c{}_{b\rangle}- 2\Pi_{c\langle a}\omega^c{}_{b\rangle}-
{2\over3}\,B^2\sigma_{ab}\,.  \label{MHDdotPi}
\end{equation}

\section{Inhomogeneous magnetised cosmologies}\label{sIMCs}
Magnetic fields appear to be everywhere in the universe. Galaxies and galactic clusters, in particular, support large-scale $B$-fields with strengths around $10^{-6}$~G and $10^{-7}$~G respectively. Although the origin of cosmic magnetism is still a mystery, it is conceivable that $B$-fields could have played some role during the process of structure formation.

\subsection{The key variables}\label{ssKVs}
The observed large-scale structure of the universe is believed to be the result of gravitational instability, a physical mechanism that allows small inhomogeneities in the density distribution of the matter to grow with time. In order to study the magnetic effects on the evolution of density perturbations, we need to define a set of key variables that describe these distortions. Following the relativistic treatments of~\cite{TB1,TB2} and~\cite{TM}, inhomogeneities in the density distribution of a magnetised medium are monitored through the dimensionless gradients
\begin{equation}
\Delta_a= {a\over\rho}\,\partial_a\rho\,, \hspace{15mm} {\rm and} \hspace{15mm} \mathcal{B}_a= {a\over B^2}\, \partial_aB^2\,.  \label{DelacBa}
\end{equation}
The former of the above describes spatial variations in the density of the baryonic matter, as measured between two neighbouring comoving observers (i.e.~flow lines), while the latter does the same for the magnetic pressure. To close the system, one also needs the auxiliary variable
\begin{equation}
\mathcal{Z}_a= a\partial_a\Theta\,,  \label{cZa}
\end{equation}
describing spatial variations in the volume expansion.

\subsection{The nonlinear equations}\label{sNEs}
The evolution of density inhomogeneities is determined by the convective derivative of $\Delta_a$. Starting from definition (\ref{DelacBa}), using the continuity equation (see expression (\ref{CEs}a)), and then applying the commutation law between  spatial gradients and convective derivatives, gives\footnote{By means of definition (\ref{dotva1}) and decomposition (\ref{pbva}), the commutation law between convective time-derivatives (i.e.~``dot-derivatives'') and spatial gradients takes the form
\begin{equation}
\partial_a\dot{\phi}= (\partial_a\phi)^{\cdot}+ {1\over3}\,\Theta\partial_a\phi+ (\sigma_{ba}+\omega_{ba})\partial^b\phi\,.  \label{cl}
\end{equation}
Note that the above (Newtonian) law also holds when $\phi$ is replaced by an arbitrary vector or tensor field.}
\begin{equation}
\dot{\Delta}_a= -\mathcal{Z}_a- (\sigma_{ba}+\omega_{ba}) \Delta^b\,.  \label{nldotDela}
\end{equation}
Unlike the relativistic case (compare to Eq.~(5.7.4) in~\cite{BMT}), matter perturbations are not coupled to those in the magnetic pressure at this stage. Instead, the coupling of the two occurs in the second-order differential equation for $\Delta_a$. The reason lies in the treatment of space and time, which in the Newtonian theory are completely separate entities.

Similarly, the nonlinear propagation formula of the expansion gradients comes from definition (\ref{cZa}). The dot-derivative of the latter, together with Raychaudhuri's equation (see (\ref{Ray})) and the commutation law (\ref{cl}) leads to
\begin{equation}
\dot{\mathcal{Z}}_a= -{2\over3}\,\Theta\mathcal{Z}_a-
{1\over2}\,\kappa\rho\Delta_a+ a\mathcal{A}_a- (\sigma_{ba}+\omega_{ba})\mathcal{Z}^b-
2a\partial_a(\sigma^2-\omega^2)\,,  \label{nldotcZa}
\end{equation}
where $\mathcal{A}_a=\partial_a(\partial^bA_b)$ by definition. Although the above contains no explicit magnetic terms, the influence of the $B$-field is incorporated within the acceleration term $\mathcal{A}_a$, as the Navier-Stokes equation ensures (see expression (\ref{NSMHD3})). Note that the relativistic analysis allows for a direct coupling between the expansion and the magnetic pressure gradients (compare to Eq.~(5.7.5) in~\cite{BMT}). There is no such coupling in (\ref{nldotcZa}) (as well as in Eq.~(\ref{nldotDela})) and the reason is partly the aforementioned difference is the treatment of space and time between the two theories, and partly the absence of a magnetic contribution to the total gravitational mass of our Newtonian system.

Finally, taking the convective derivative of (\ref{DelacBa}b), using the conservation law of the magnetic pressure (see Eq.~(\ref{MHDdotB2})), the nonlinear commutation law (\ref{cl}) and expression (\ref{nldotDela}), we obtain
\begin{eqnarray}
\dot{\mathcal{B}}_a&=& {4\over3}\,\dot{\Delta}_a- (\sigma_{ba}+\omega_{ba}) \left(\mathcal{B}^b-{4\over3}\,\Delta^b\right)- {2a\over B^2}\, \partial_a(\sigma_{bc}\Pi^{bc})+ {2\over B^2}\, \sigma_{bc}\Pi^{bc}\mathcal{B}_a\,.  \label{nldotcBa}
\end{eqnarray}
The above monitors spatial variations in the (isotropic) magnetic pressure, between a pair of neighbouring observers. We point out the direct connection between the magnetic and the matter inhomogeneities, reflected in the first two terms of Eq.~(\ref{nldotcBa}). This coupling, which also occurs in relativistic studies -- compare the above to the propagation formula (5.7.6) in~\cite{BMT}, will prove particularly useful at the linear level.

\section{The linear regime}\label{sLR}
The set (\ref{nldotDela})-(\ref{nldotcBa}) governs the nonlinear evolution of density inhomogeneities in a magnetised Newtonian medium of high electrical conductivity and zero total charge. In this section we will linearise these expressions around the Newtonian analogue of the Einstein-de Sitter universe.

\subsection{The unperturbed background}\label{ssUB}
Let us assume a homogeneous and isotropic background universe filled with a barotropic perfect fluid (with $p=p(\rho)$ and $\pi_{ab}=0$) and allow for a weak and completely random (i.e.~statistically homogeneous and isotropic) magnetic field (see also~\cite{BMT}). Later we will also consider a pure (i.e.~a magnetic-free) Einstein-de Sitter unperturbed model (see Appendix~B). Assuming a random $B$-field means that $B_a=0$ on average in the background. On the other hand, the magnetic pressure does not average to zero (i.e.~$B^2\neq0$ on average) and has a time dependence only. The randomness of the $B$-field guarantees that its presence does not destroy the uniformity of our zero-order model, although it adds to the total pressure of the system. In addition, the overall weakness of the field ensures that the Alfv\'en speed (defined by $c_a^2=B^2/\rho$) is always well below unity.\footnote{Demanding that $c_a^2\ll1$ places no real constraint on the magnetic field. For instance, setting $\rho\simeq10^{-47}~{\rm GeV}^{4}$, which is the critical density of the universe today, translates into $B\ll10^{-4}$~G at present. The latter condition is satisfied by all the known large-scale magnetic fields.} Thus, ignoring the cosmological constant, our unperturbed model is governed by the set\footnote{Hereafter, barred variables will indicate background quantities of zero perturbative order.}
\begin{equation}
\dot{\bar{\rho}}= -3H\bar{\rho}\,, \hspace{15mm}
H^2= {1\over3}\,\kappa\bar{\rho}\,, \hspace{15mm} \dot{H}= -{1\over2}\,\kappa\bar{\rho}  \label{FRW}
\end{equation}
and
\begin{equation}
\left(\bar{B}^2\right)^{\cdot}= -4H\bar{B}^2\,,  \label{bdotB2}
\end{equation}
since $\bar{\Theta}=3H=3\dot{a}/a$ to zero perturbative order. Solving (\ref{FRW}b) for the scale factor, leads to $a\propto t^{2/3}$, which is the familiar evolution law of the Einstein-de Sitter universe. Also, following (\ref{FRW}a) and (\ref{bdotB2}), gives $\bar{\rho}\propto a^{-3}$ and $\bar{B}^2\propto a^{-4}$, guaranteeing that the magnetic pressure decays faster than the density of the matter. Finally, it is straightforward to show that
\begin{equation}
\left(c_a^2\right)^{\cdot}= -Hc_a^2\,,  \label{Alfven}
\end{equation}
to zero order. The above implies that $c_a^2\propto a^{-1}\propto t^{-2/3}$ to ensure that the Alfv\'en speed decreases inversely proportionally with the dimensions of the universe.

\subsection{Linear magnetised perturbations}\label{ssLMPs}
The background model determines our zero-order solution. Perturbatively speaking, quantities with nonzero background value will comprise our zero-order variables, while those vanishing there will be treated as first-order perturbations. When linearising, terms of perturbative order higher than the first will be neglected. Thus, given our background, the only zero-order quantities are the Hubble parameter ($H$), the density of the matter ($\rho$) and the magnetic pressure ($B^2$). All the rest are of order one perturbatively. Note that differentiation (temporal or spatial) leaves the order of a perturbed quantity unaffected.

Applying the above described linearisation scheme to our basic equations, namely to the nonlinear formulae (\ref{nldotDela})-(\ref{nldotcBa}), the latter reduce to
\begin{equation}
\dot{\Delta}_a= -\mathcal{Z}_a\,,  \label{ldotDela}
\end{equation}
\begin{equation}
\dot{\mathcal{Z}}_a= -2H\mathcal{Z}_a-
{1\over2}\,\kappa\bar{\rho}\Delta_a+ a\mathcal{A}_a  \label{ldotcZa}
\end{equation}
and
\begin{equation}
\dot{\mathcal{B}}_a= {4\over3}\,\dot{\Delta}_a\,,  \label{ldotcBa}
\end{equation}
respectively. In contrast with the relativistic treatment (e.g.~see \S~6.3.4 in~\cite{BMT} for comparison), the magnetic effects on linear density perturbations propagate only through the acceleration term on the right-hand side of Eq.~(\ref{ldotcZa}). Recalling that $\mathcal{A}_a= \partial_a(\partial^bA_b)$, assuming an ideal fluid (i.e.~setting $\pi_{ab}=0$ in (\ref{NSMHD3})) and using definitions (\ref{DelacBa}) we obtain
\begin{equation}
a\mathcal{A}_a= -c_s^2\partial^2\Delta_a- {1\over2}\,c_a^2\partial^2\mathcal{B}_a+ {2a\over\bar{\rho}}\,\partial_a\left(\sigma_B^2-\omega_B^2\right)\,,  \label{lcAterm1}
\end{equation}
to linear order.\footnote{Perturbatively, the gradient $\partial_aB^2$ is treated as a first-order quantity, since it vanishes in the background. Then, given that $\partial_aB^2=2B^b\partial_aB_b$, both $B_a$ and $\partial_bB_a$ are of perturbative order half. This makes $\sigma_B^2$ and $\omega_B^2$ of order one.} Note that $c_s^2=\dot{\bar{p}}/\dot{\bar{\rho}}\ll1$ defines the unperturbed adiabatic sound speed and $c_a^2=\bar{B}^2/\bar{\rho}$ (with $c_a^2\ll1$ at all times) the background Alf\'en speed. Also, $\sigma_B^2=\partial_{\langle b}B_{a\rangle}\partial^{\langle b}B^{a\rangle}/2$ and $\omega_B^2= \partial_{[b}B_{a]}\partial^{[b}B^{a]}/2$ by construction. These variables may be seen as the magnetic analogues of the shear and of the vorticity respectively. Finally, following Eq.~(\ref{ldotcBa}), the linearised gradients of the magnetic pressure evolve in tune with those in the density of the matter. More specifically
\begin{equation}
\mathcal{B}_a= {4\over3}\,\Delta_a+ \mathcal{C}\,,  \label{cBa}
\end{equation}
with $\mathcal{C}$ representing the integration constant. The system (\ref{ldotDela})-(\ref{ldotcBa}) monitors the linear evolution of density inhomogeneities in a weakly magnetised Newtonian medium of zero electrical resistivity. Alternatively, one can take the convective derivative of (\ref{ldotDela}) and then employ Eqs. (\ref{ldotDela}), (\ref{ldotcZa}) to eliminate $\mathcal{Z}_a$. Finally. using the auxiliary expression (\ref{lcAterm1}) and the linear result (\ref{cBa}), the aforementioned set is replaced by
\begin{equation}
\ddot{\Delta}_a= -2H\dot{\Delta}_a+ {1\over2}\,\kappa\bar{\rho}\Delta_a+ \left(c_s^2+{2\over3}\,c_a^2\right)\partial^2\Delta_a- {2a\over\bar{\rho}}\,\partial_a\left(\sigma_B^2-\omega_B^2\right)\,,  \label{lDelddot}
\end{equation}
which is a wave-like differential equation with additional terms due to the universal expansion, the presence of matter and the action of the magnetic field.

\subsection{The magnetic effects}\label{ssMEs}
To linear order, the effects of the $B$-field on density perturbations propagate via the last two terms on the right-hand side of Eq.~(\ref{lDelddot}). The former is due to the (positive) magnetic pressure and the latter is the result of the magnetic tension. Recall that the tension reflects the reaction of the field lines against any agent distorting their equilibrium. As a result, the tension effects are expected to become more prominent when the entanglement of the field lines is sufficiently large~\citep{TS}. Put another way, although the magnetic shear and vorticity stresses (i.e.~the terms $\sigma_B^2= \partial_{\langle b}B_{a\rangle}\partial^{\langle b}B^{a\rangle}/2$ and $\omega_B^2=\partial_{[b}B_{a]} \partial^{[b}B^{a]}/2$ respectively) are first-order perturbatively, they are expected to play a secondary role, unless the magnetic configuration is highly distorted. Thus, ignoring (momentarily only -- see Appendix~A) these two terms and then harmonically decomposing the remaining expression gives\footnote{We obtained (\ref{lDelddot1}) by setting $\Delta_a=\Sigma_n\Delta_{(n)}\mathcal{Q}^{(n)}_a$, with $\mathcal{Q}_a^{(n)}$ representing standard vector harmonic functions and $n>0$ the comoving wavenumber of the associated mode. Also, by construction, $\partial_a\Delta_{(n)}=0 =\dot{\mathcal{Q}}^{(n)}_a$ and $\partial^2\mathcal{Q}^{(n)}_a= -(n/a)^2\mathcal{Q}^{(n)}_a$ (e.g.~see~\cite{E3,TCM}).}
\begin{equation}
\ddot{\Delta}_{(n)}= -2H\dot{\Delta}_{(n)}+ {1\over2}\,\kappa\bar{\rho}\left[1-{2\over3}\, \left(c_s^2+{2\over3}\,c_a^2\right) \left({\lambda_H\over\lambda_n}\right)^2\right]\Delta_{(n)}\,,  \label{lDelddot1}
\end{equation}
where $\lambda_H=1/H$ is the Hubble radius and $\lambda_n=a/n$ is the scale of the inhomogeneity. Note that, we have arrived to a homogeneous differential equation, as opposed to an inhomogeneous one (e.g.~see~\cite{KOR,TS}).\footnote{A homogeneous differential equation, but of order four, was also obtained by~\cite{RR}.} This is also true when the tension stresses have been incorporated into the system (see Appendix~A below). As a result, extracting analytical solutions becomes a relatively straightforward process, which is an attractive feature of the covariant formalism.

It is worth pointing out that one can use Eq.~(\ref{lDelddot1}) even when the universe is dominated by pressureless non-baryonic cold dark matter (CDM), provided there is no relative velocity between the baryonic and the dark components. This happens because the magnetic field does not interact with the CDM, which contributes only to the total gravitational field through Raychaudhuri's formula (see~Eq.~(\ref{Ray}) in \S~\ref{ssHDs}). Therefore, qualitatively speaking, the only change would be in the background density (i.e.~in $\bar{\rho}$ -- see expression (\ref{lDelddot1})), which would now correspond to the dark sector. Nevertheless, the background dynamics remain unaffected, since both the baryonic and the dark-matter densities obey the same evolution law.\footnote{The role of CDM should become more prominent at the nonlinear level, when the baryonic matter starts falling into the `potential wells' of the dark matter. This increases the strength of the magnetic field, which remains frozen-in with the baryons, beyond the adiabatic limit~\citep{BrMT,DBL}. The enhanced $B$-field could then backreact and have a stronger effect on the evolution of density perturbations.} On the other hand, the background dynamics will change drastically if we assume that the universe is dominated by dark energy or by a positive cosmological constant. Consequently, given that current observations support the idea of a recent universal acceleration, Eq.~(\ref{lDelddot1}) holds from the time of recombination (at $z\simeq10^3$) until the start of the accelerated phase (at $z\simeq0.5$).

An additional issue is the working domain of expression (\ref{lDelddot1}) and the spectrum of the background magnetic field. Since we use Newtonian gravity, we can only consider subhorizon-sized perturbations (i.e.~those with $\lambda_n<\lambda_H$). Also, if one wants to focus on the magnetic effects and bypass those of the baryonic pressure, as we are going to do next, the scales of interest should not be very small and probably not much smaller than those of a galactic cluster. On all these lengths, the spectrum of the background (the random) magnetic field has been treated as scale-invariant.

Expression (\ref{lDelddot1}) describes magnetosonic waves, propagating with an effective ``sound'' speed equal to $c_{ms}^2=c_s^2+(2/3)c_a^2$ (see also Eq.~(\ref{lDelddot})). Consequently, the $B$-field adds to the total pressure of the system, which in turn increases the associated Jeans length. In fact, even when dealing with pressure-free matter, there is a purely magnetic Jeans length given by
\begin{equation}
\lambda_J= {2\over3}\,c_a\lambda_H\,.  \label{BJeans}
\end{equation}
The latter agrees with the result of the relativistic treatment (e.g.~see \S~7.4.1 in~\cite{BMT}), as well as with that of the Newtonian studies (e.g.~see~\cite{KOR,TS}). Note that $c_a\propto a^{-1/2}$ and $\lambda_H\propto a^{3/2}$, which imply that $\lambda_J\propto a\propto t^{3/2}$. In other words, the physical size of the magnetic Jeans length increases in tune with the dimensions of the universe. Consequently, when the first stars where formed, between $z\simeq15$ and $z\simeq20$ (e.g.~see~\cite{TS}), the magnetic Jeans length (and the physical size of the first structures allowed to collapse gravitationally) was approximately 15 to 20 times smaller than it is today. Similarly, at the start of the accelerated phase (at $z\simeq0.5$) the magnetic Jeans length was slightly (roughly 1.5 times) smaller than its current value.

Using definition (\ref{BJeans}) one can estimate the scale of the magnetic Jeans length for some characteristic values of the background $B$-field. The same length-scale also provides the size of the first gravitationally-bound structures to form, in a scenario where the only pressure support comes from cosmological magnetic fields. For instance, setting $B\sim10^{-7}$~G, which is the typical strength of the cluster magnetic fields and also the maximum allowed by primordial nucleosynthesis, gives $B^2\sim10^{-54}~{\rm GeV}^4$. Then, if we set the background density equal to the critical density (i.e.~for $\bar{\rho}=\rho_c\sim10^{-47}~{\rm GeV}^4$, we find that $\lambda_J\sim1$~Mpc. The latter is the size of a small galactic cluster. Alternatively, for $B\sim10^{-9}$~G, which is the upper limit of a homogeneous $B$-field allowed by the Cosmic Microwave Background (CMB), the magnetic Jeans length drops to $\lambda_J\sim10$~Kpc. Finally, recent surveys have suggested the presence of intergalactic magnetic fields close to $10^{-15}$~G~\citep{TGFBGC,AK,NV}. Adopting this value, expression (\ref{BJeans}) gives $\lambda_J\sim10^{-2}$~pc. Note that, as we have pointed out earlier, our analysis holds until the onset of the universal acceleration (at $z\simeq0.5$), at which moment the above given values are slightly smaller (by roughly 1.5 times -- see earlier discussion). Overall, assuming that cosmological magnetic fields are the only sources of pressure support after recombination and given the presumed weakness of such fields, the first structures to form will have typical sizes of a small galactic cluster or smaller (see also~\cite{KOR}). In general, the $B$-field adds to the ambient thermal pressure of the matter and thus increases the effective sound-speed and the associated Jeans length by a small amount.

Ignoring the matter pressure means that Eq.~(\ref{lDelddot1}) can be solved analytically. In particular, given that $H=2/3t$ and $\kappa\bar{\rho}=4/3t^2$ (for both baryonic dust and CDM), the aforementioned differential equation reads
\begin{equation}
{{\rm d}^2\Delta_{(n)}\over{\rm d}t^2}= -{4\over3t}{{\rm d}{\Delta}_{(n)}\over{\rm d}t}+ {2\over3t^2}\left[1-{4\over9}\,c_a^2 \left({\lambda_H\over\lambda_n}\right)^2\right]\Delta_{(n)}\,.  \label{lDelddot2}
\end{equation}
The latter, which on subhorizon scales and in the absence of curvature effects agrees fully with its relativistic analogue (e.g.~see \S~7.4.1 in~\cite{BMT}), accepts the solution
\begin{equation}
\Delta_{(n)}= \mathcal{C}_1\,t^{{1\over6}\,(\sqrt{25-24\alpha}-1)}+ \mathcal{C}_2\,t^{-{1\over6}\,(\sqrt{25-24\alpha}+1)}\,,  \label{lDel1}
\end{equation}
with
\begin{equation}
\alpha= {4\over9}\,c_a^2 \left({\lambda_H\over\lambda_n}\right)^2= \left({\lambda_J\over\lambda_n}\right)^2\,,  \label{alpha}
\end{equation}
being a positive constant (recall that $c_a^2=B^2/\rho\propto a^{-1}$ and $\lambda_H/\lambda_n\propto a^{1/2}$). The above monitors the effects of cosmological $B$-fields on linear inhomogeneities in the density of baryonic dust. Note that, given the weakness of the magnetic fields (i.e.~since $c_a^2=B^2/\rho\ll1$), the typical values of the $\alpha$ are smaller than unity. It is conceivable, however that on sufficiently small wavelengths (i.e.~those with $(\lambda_H/\lambda_n)^2>(75/32)c_a^{-2}$), the value of $\alpha$ may exceed unity (see third case next). Having said that, we should also stress that on very small scales the nonlinear effects start becoming important and our linear approximation is expected to break down.

The first point to emphasise is that when there is no  $B$-field (i.e.~for $\alpha=0$), solution (\ref{lDel1}) reduces to its non-magnetised (dust) counterpart, where $\Delta_{(n)}=\mathcal{C}_1t^{2/3}+ \mathcal{C}_2t^{-1}$ on all scales. Otherwise, the magnetic effect depends on the value of the $\alpha$, namely on the strength of the field and on the scale of the perturbed mode (see definition (\ref{alpha}) above). We may therefore distinguish between three alternative cases:

\begin{itemize}
\item When $0<\alpha<1$, one can easily show that $\lambda_n>\lambda_J$ and $0<(\sqrt{25-24\alpha}-1)/6<2/3$. Therefore, on scales larger than the magnetic Jeans length, solution (\ref{lDel1}) reduces to a power-law that contains one growing mode. The latter, however, increases slower than in non-magnetised cosmologies.
\item Assuming that $1\leq\alpha\leq25/24$, namely that $\sqrt{24/25}\lambda_J\leq\lambda_n\leq\lambda_J$, gives $-1/6\leq(\sqrt{25-24\alpha}-1)/6\leq0$. Within this narrow margin of wavelengths, density perturbations either remain constant or decay at a pace no faster than $\Delta_{(n)}\propto t^{-1/6}$. It looks like a 'transition` range, where the gravitational pull is not strong enough to make the perturbations grow and the support of the field's pressure is unable to make them oscillate (see next case). To the best of our knowledge, this type of behaviour has not been noted before. If observed, it could in principle provide evidence of a magnetic presence on relatively large scales. It is conceivable, however, that other sources of pressure could mimic this magnetic effect.
\item Finally, for $\alpha>25/24$ we have $\lambda_n<\sqrt{24/25}\lambda_J$ and $25-24\alpha<0$. Then, setting $\sqrt{25-24\alpha}=\imath\sqrt{24\alpha-25}$ in the right-hand side of (\ref{lDel1}), shows that the inhomogeneity oscillates with decreasing amplitude. More specifically, when $\alpha>25/24$, solution (\ref{lDel1}) takes the oscillatory form
\begin{equation}
\Delta_{(n)}= t^{-1/6}\left[\mathcal{C}_1\cos\left({1\over6}\sqrt{24\alpha-25}\ln t\right)+\mathcal{C}_2\sin\left({1\over6}\sqrt{24\alpha-25}\ln t\right)\right]\,,  \label{lDel2}
\end{equation}
while its amplitude drops as $t^{-1/6}$. As we mentioned above, solution (\ref{lDel2}) holds on small scales (i.e.~those with $(\lambda_H/\lambda_n)^2> (75/32)c_a^{-2}\Leftrightarrow\lambda_n<\sqrt{24/25}\lambda_J$), where the magnetic pressure is strong enough to support against the gravitational pull of the matter.
\end{itemize}

In summary, one could argue that the $B$-field has a negative effect on the growth of linear density inhomogeneities, since the latter either grow slower or decay with time. Analogous (negative) effects were observed both in relativistic studies (e.g.~see~\cite{TB2} for the radiation era), as well as in Newtonian treatments (e.g.~see~\cite{RR}). In all cases, the reason is the extra pressure that the field introduces into the system and the overall magnetic effect is proportional to its strength. Although, the latter drops with time, the overall effect is decided by the $\alpha$-parameter which is scale dependent (see definition (\ref{alpha})) and remains constant.

\section{Discussion}\label{sD}
The continuous detection of large-scale magnetic fields in the universe, especially those found in galaxy clusters and in distant proto-galactic formations, adds support to the idea of primordial magnetism. Assuming that $B$-fields have cosmological origin, it is conceivable that they could have played some role during the process of structure formation. In the present article we have re-considered the implications of large-scale magnetic fields for the linear evolution of density perturbations, in the baryonic component of the matter, after recombination. This allowed us to employ a Newtonian approach, which is believed to provide an adequate treatment of the problem in hand on subhorizon scales and once the universe has gone past equipartition and decoupling. Technically, we proceeded by adopting the Newtonian version of the 1+3 covariant approach to cosmology, which also facilitated the direct comparison of our study with the previous relativistic treatments. In addition, one should in principle be able to exploit the techniques developed here, as well as the experience gained, to improve the general relativistic study of the issue. For instance, the technique used to incorporate the tension effects of the $B$-field and still arrive at a homogeneous differential equation for the density perturbation (see Appendix~A), is likely to work within the relativistic framework as well.

We looked at the problem by adopting two alternative scenarios. The first allowed for a weak and fully random (statistically homogeneous and isotropic) $B$-field in our Einstein-de Sitter background, while the second employed a pure (magnetic-free) unperturbed model. In the latter case the field did not have any practical effect on the linear evolution of density inhomogeneities, since it simply added a constant mode in between the growing and the decaying modes of the non-magnetised study (see Appendix~B). Introducing a weak magnetic field into the background, on the other hand, strengthened its overall impact and lead to explicit magnetic terms in the solutions. The latter were found to differ qualitatively, depending on the strength of the $B$-field and on the scale of the perturbed mode. Roughly speaking, on wavelengths longer than the so-called magnetic Jeans length, density perturbations grow but at a pace slower than in the magnetic-free models. Inhomogeneities spanning scales smaller than the aforementioned Jeans length, however, oscillate with time-decreasing amplitude. There is also a narrow window of wavelengths, just inside the magnetic Jeans length, where the perturbations simply decay with time. Finally, incorporating the effects of the magnetic tension did not seem to cause any significant change to our results (see Appendix~A). Practically speaking, whether the aforementioned effects could leave an observable signature depends on the actual strength of the magnetic fields and on the sensitivity of the instruments. Qualitatively speaking, we expect to see an overall `negative' effect, since the field's presence either slows down the standard growth-rate of the inhomogeneities, or forces them to decay. Analogous (negative) effects were also observed in the relativistic study of the radiation era and in all cases the reason is the extra pressure that the $B$-field introduces into the system.

It is also worth mentioning that our Newtonian analysis also confirmed the existence of a purely magnetic Jeans length, the scale of which depends on the strength of the field. This characteristic length-scale determines the size of the first structures to form gravitationally, assuming that magnetic fields are the only sources of pressure support in the post-recombination universe. Given the presumed weakness of cosmological magnetic fields, we expect that the aforementioned first structures are small, with their typical sizes never exceeding that of a small galactic cluster. Generally speaking, the $B$-field adds to the ambient thermal pressure of the matter and thus increases the total effective Jeans length by a small amount.

\appendix

\section{Incorporating the magnetic tension}\label{sA1}
The magnetic effects analysed in section \S~\ref{ssMEs} account for the (positive) pressure of the field only. To incorporate the tension component of the Lorentz force (see decomposition (\ref{NSMHD3}) in \S~\ref{ssIMHD}), one needs to include the magnetic shear and the magnetic vorticity terms ($\sigma_B^2$ and $\omega_B^2$ respectively) on the right-hand side of Eq.~(\ref{lDelddot}). Taking the time-derivative of the latter, using a set of auxiliary relations (see Eqs.~(\ref{aux1}) and (\ref{aux2}) below) and then harmonically decomposing the resulting expression, gives
\begin{equation}
{{\rm d}^3\Delta_{(n)}\over{\rm d}t^3}= -{10\over3t}\,{{\rm d}^2\Delta_{(n)}\over{\rm d}t^2}- {2(1+\alpha)\over3t^2}\, {{\rm d}\Delta_{(n)}\over{\rm d}t}\,,  \label{lDeldddot1}
\end{equation}
where $\alpha=(\lambda_J/\lambda_n)^2=$~constant (see definition (\ref{alpha})). The above solves to give
\begin{equation}
\Delta_{(n)}= \mathcal{C}_1\,t^{{1\over6}\,(\sqrt{25-24\alpha}-1)}+ \mathcal{C}_2\,t^{-{1\over6}\,(\sqrt{25-24\alpha}+1)}+ \mathcal{C}_3\,,  \label{lDel3}
\end{equation}
which is identical to solution (\ref{lDel1}) with the exception of an additional constant mode on its right-hand side. The latter carries the effects of the magnetic tension and plays a role only when $\alpha>25/24$, in which case the other two modes decay with time (see also \S~\ref{ssMEs} above). Overall, when dealing with linear density perturbations, the magnetic tension is typically a secondary player (see~\cite{BMT} for a discussion of the relativistic case). On the other hand, the tension plays the prominent role when studying the magnetic effects on linear rotational distortions, both in the Newtonian and the relativistic framework (see~\cite{W,DT}).

Technically speaking, one arrives at Eq.~(\ref{lDeldddot1}) by using the background relations (\ref{FRW}a), (\ref{FRW}c) and (\ref{Alfven}), together with the auxiliary linear evolution laws
\begin{equation}
\left(\partial_a\sigma_B^2\right)^{\cdot}= -7H\partial_a\sigma_B^2\,, \hspace{10mm} \left(\partial_a\omega_B^2\right)^{\cdot}= -7H\partial_a\omega_B^2  \label{aux1}
\end{equation}
and
\begin{equation}
\left(\partial^2\Delta_a\right)^{\cdot}= \partial^2\dot{\Delta}_a- 2H\partial^2\Delta_a\,,  \label{aux2}
\end{equation}
which reflect the non-commutativity between convective and time derivatives (see expression (\ref{cl}) in footnote~3). Note that the last two relations stem from the fact that $(\partial_bB_a)^{\cdot}=-3H\partial_bB_a$ to first order (see also Appendix~C in~\cite{DdSTB}). Finally, we have obtained the homogeneous differential equation (\ref{lDeldddot1}), by going back to Eq.~(\ref{lDelddot}) and using it to eliminate the magnetic shear and vorticity terms.

\section{The case of a magnetic-free background}\label{sA2}
So far, we have analysed the magnetic effects on the evolution of linear density perturbation, assuming a (Newtonian) Einstein-de Sitter background that contains a weak, random (i.e.~statistically homogeneous and isotropic) $B$-field. Here, primarily for comparison reasons, we will consider a pure (i.e.~a magnetic-free) background cosmology. In the absence of the field, the zero-order equations are given by the set (\ref{FRW}a)-(\ref{FRW}c). Then, treating the magnetic pressure ($B^2$) as a first-order quantity, the key linear variables are
\begin{equation}
\Delta_a= {a\over\bar{\rho}}\,\partial_a\rho\,, \hspace{10mm} Z_a= a\partial_a\Theta \hspace{10mm} {\rm and} \hspace{10mm} \mathcal{B}_a= {a\over\bar{\rho}}\,\partial_aB^2\,,
\end{equation}
where $\mathcal{B}_a$ has now been normalised using the background density of the matter (compare to definition (\ref{DelacBa}b) in \S~\ref{ssKVs}). For simplicity, let us set the fluid pressure to zero and ignore the magnetic shear and vorticity terms. Under these assumptions, the above defined variables evolve according to the linear formulae
\begin{equation}
\dot{\Delta}_a= -Z_a\,, \hspace{10mm} \dot{Z}_a= -2HZ_a- {1\over2}\,\kappa\bar{\rho}\Delta_a- {1\over2}\,\partial^2\mathcal{B}_a  \label{lDelaZacBa1}
\end{equation}
and
\begin{equation}
\dot{\mathcal{B}}_a= -H\mathcal{B}_a\,, \label{lDelaZacBa2}
\end{equation}
respectively. Proceeding in a way analogous to that outlined in Appendix~A, the above expressions combine to the differential equation
\begin{equation}
{{\rm d}^3\Delta_{(n)}\over{\rm d}t^3}= -{10\over3t}\,{{\rm d}^2\Delta_{(n)}\over{\rm d}t^2}- {2\over3t^2}\, {{\rm d}\Delta_{(n)}\over{\rm d}t}\,,  \label{lDeldddot2}
\end{equation}
which accepts the solution
\begin{equation}
\Delta_{(n)}= \mathcal{C}_1\,t^{2/3}+ \mathcal{C}_2\,t^{-1}+ \mathcal{C}_3\,.  \label{lDel4}
\end{equation}
The main feature here is the absence of explicit magnetic terms on the right-hand side (compare to solutions (\ref{lDel1}), (\ref{lDel2}) and (\ref{lDel3}) of the weakly magnetised background). Instead, the $B$-field has simply added a (physically unimportant) constant mode in between the growing and the decaying modes of the non-magnetised case. Therefore, removing the magnetic field from the background model has weakened its overall effect, something clearly reflected in the above solution. Put another way, incorporating a weak (fully random) $B$-field in the background enhances its input by a small amount, which however is enough to allow for explicit magnetic terms in the solutions (see (\ref{lDel1}) and (\ref{lDel2})). The latter clarify qualitatively the role of the field and also help to quantify its impact.

\end{document}